\documentclass[sigconf]{acmart}
\AtBeginDocument{%
  \providecommand\BibTeX{{%
    \normalfont B\kern-0.5em{\scshape i\kern-0.25em b}\kern-0.8em\TeX}}}

\usepackage{float}

\usepackage{wrapfig}
\usepackage{todonotes}

\usepackage{listings}

\definecolor{codegreen}{rgb}{0,0.6,0}
\definecolor{codegray}{rgb}{0.5,0.5,0.5}
\definecolor{codepurple}{rgb}{0.58,0,0.82}  
\definecolor{backcolour}{rgb}{1,1,1}

\def\beskow{Beskow}
\def\kebne{Kebnekaise}

\usepackage[compact]{titlesec}

\titlespacing{\section}{0pt}{*0}{*0}
\titlespacing{\subsection}{0pt}{*0}{*0}
\titlespacing{\subsubsection}{0pt}{*0}{*0}

\usepackage[ruled,linesnumbered]{algorithm2e}
\usepackage{amsmath}
\usepackage[htt]{hyphenat}

\newcommand{\tens}[1]{%
  \mathbin{\mathop{\otimes}\limits_{#1}}%
}
\AtBeginDocument{%
  \providecommand\BibTeX{{%
    \normalfont B\kern-0.5em{\scshape i\kern-0.25em b}\kern-0.8em\TeX}}}
\usepackage{hyperref}
\usepackage{capt-of}
\usepackage{booktabs}
\usepackage{varwidth}
\usepackage{cleveref}
\usepackage[T1]{fontenc}
\usepackage[font=small,labelfont=bf,tableposition=top]{caption}
\DeclareCaptionLabelFormat{andtable}{#1~#2  \&  \tablename~\thetable}

\setcopyright{acmcopyright}
\copyrightyear{2021}
\acmYear{2021}
\acmDOI{10.1145/1122445.1122456}

\acmConference[HEART'21]{HEART'21: ACM International Symposium on Highly Efficient Accelerators and Reconfigurable Technologies}{June 21--23, 2021}{Virtual Event}
\acmBooktitle{HEART'21: ACM International Symposium on Highly Efficient Accelerators and Reconfigurable Technologies, 2021}
\acmPrice{15.00}
\acmISBN{978-1-4503-XXXX-X/18/06}

\begin{document}

\title{StreamBrain: An HPC Framework for Brain-like Neural Networks on CPUs, GPUs and FPGAs}


\author{Artur Podobas$^1$, Martin Svedin$^1$, Steven W. D. Chien$^1$, Ivy B. Peng$^3$, Naresh Balaji Ravichandran$^1$, Pawel Herman$^1$, Anders Lansner$^{1,2}$, and Stefano Markidis$^1$}
\affiliation{
\institution{$^1$ KTH Royal Institute of Technology, Stockholm, Sweden}
}
\affiliation{
\institution{$^2$ Stockholm University, Stockholm, Sweden}
}
\affiliation{
\institution{$^3$ Lawrence Livermore National Laboratory, CA, USA}
}

\copyrightyear{2021}
\acmYear{2021}
\setcopyright{usgovmixed}\acmConference[HEART '21]{International Symposium on
Highly Efficient Accelerators and Reconfigurable Technologies}{June 21--23,
2021}{Online, Germany}
\acmBooktitle{International Symposium on Highly Efficient Accelerators and
Reconfigurable Technologies (HEART '21), June 21--23, 2021, Online, Germany}
\acmPrice{15.00}
\acmDOI{10.1145/3468044.3468052}
\acmISBN{978-1-4503-8549-7/21/06}

\renewcommand{\shortauthors}{Podobas, et al.}

\begin{abstract}
The modern deep learning method based on backpropagation has surged in popularity and has been used in multiple domains and application areas. At the same time, there are other -- less-known -- machine learning algorithms with a mature and solid theoretical foundation whose performance remains unexplored. One such example is the brain-like Bayesian Confidence Propagation Neural Network (BCPNN). In this paper, we introduce StreamBrain-- a framework that allows neural networks based on BCPNN to be practically deployed in High-Performance Computing systems. StreamBrain is a domain-specific language (DSL), similar in concept to existing machine learning (ML) frameworks, and supports backends for CPUs, GPUs, and even FPGAs. We empirically demonstrate that StreamBrain can train the well-known ML benchmark dataset MNIST within seconds, and we are the first to demonstrate BCPNN on STL-10 size networks. We also show how StreamBrain can be used to train with custom floating-point formats and illustrate the impact of using different bfloat variations on BCPNN using FPGAs.
\end{abstract}




\settopmatter{printacmref=false} 
\renewcommand\footnotetextcopyrightpermission[1]{} 
\pagestyle{plain} 

\settopmatter{printacmref=false}
\maketitle

\section{Introduction}

The recent surge in popularity of Artificial Neural Networks (ANNs) is attributed to how they effectively map to existing high-performance hardware.  Most deep learning~\cite{lecun2015deep} frameworks (e.g., Keras~\cite{gulli2017deep}) are implemented as (or transformed into) a series of dense matrix multiplications (GEMM). Dense matrix multiplication, coincidentally, has been the prime computation that has driven the assessment of High-Performance Computing (HPC) for the past decades through the TOP500 (\texttt{https://top500.org/}) performance assessment project. The famed neural network AlexNET~\cite{krizhevsky2012imagenet} was realized by using Graphics Processing Units (GPUs) optimized for dense matrix multiplications, ultimately sparking the renewed interest in many-layer ANNs that we enjoy today. At the same time, there are a large number of alternative neural network models that host a sound and solid theoretical base, but which, to this day, remains empirically untested on a large scale.  One such model is the Bayesian Confidence Propagation Neural Network (BCPNN)~\cite{johansson2007towards,ravichandran2020learning}.

BCPNN is a brain-like neural network, that builds on a Hebbian-like learning principle derived from Bayes theorem. The main building blocks are so-called hypercolumns units (HCUs) (considered to be the building block of the human cortex~\cite{mountcastle1997columnar}), which account for computations in local receptive fields (e.g., share input pixels of an image). BCPNN can be used to implement unsupervised, semi-supervised, and supervised learning.  It also features runtime-adaptable structural plasticity, which facilitates remapping of cortical components such as the connectivity of HCUs to maximize information entropy of the system, which can lead to better features~\cite{ravichandran2020learning}. BCPNN has recently been used for synaptic plasticity in large scale spiking cortex models of working memory function~\cite{fiebig2020indexing,fiebig2017spiking}  and temporal sequence learning and generation, implemented on SpiNNaker~\cite{knight2016large}.  BCPNN was further shown to reach 98.58\% accuracy~\cite{ravichandran2020brain} on the famed MNIST~\cite{lecun1998mnist} classification benchmark. This is lower than reached by supervised gradient descent methods, but comparable to other methods using the unsupervised generation of hidden representation~~\cite{ravichandran2020brain}. BCPNN has also been considered for ASIC acceleration~\cite{stathis2020ebrainii}. Unfortunately, despite the solid theory that BCPNN holds, it still remains non-trivial for non-experts to adapt and explore the system.

In this work, we propose a high-level domain-specific language (DSL)/API to interface and use BCPNN for practical deployment in future supercomputers or data-centers. \textit{Contrary to prior BCPNN work (which focuses on its theory), the present paper focuses on how to map BCPNN in order to leverage modern state-of-the-art supercomputing resources and accelerators.} We create a domain-specific language -- conceptually similar to that of Keras~\cite{gulli2017deep} -- that hosts functionality for creating, training, and inferring BCPNN-based neural networks. Our implementation, called \texttt{StreamBrain}, supports device heterogeneity, including \textbf{(i)} an OpenMP- or MPI-based~\cite{chandra2001parallel}, hand-vectorized general-purpose version, \textbf{(ii)} a highly-parallel GPU version based on the Compute Unified Device Architecture (CUDA), and \textbf{(iii)} a prototype for a Field-Programmable Gate Arrays (FPGA) version based on OpenCL~\cite{czajkowski2012opencl} and High-Level Synthesis (HLS) on Intel devices, capable of using variable-precision numerical formats. 

We claim the following three contributions: \textbf{(i)} StreamBrain, a Keras-inspired~\cite{gulli2017deep} DSL for implementation, evaluation, and deployment of BCPNN for use in future high-performance computers and data-centers, \textbf{(ii)} Analysis, implementation, validation, and empirical evaluation of three different BCPNN backends for CPUs, GPUs, and FPGAs, on the MNIST and STL-10 benchmarks and on two modern supercomputers, \textbf{(iii)} Empirical evaluation on both batching and variable-precision arithmetics on the BCPNN model.


\section{The BCPNN Model}\label{sec:BCPNN}
BCPNN is a brain-like neural network model that has both an abstract rate-based formulation and detailed spiking neuron-based formulation. In this paper, we focus on the rate-based formulation. We model the neural network problem with a collection of random variables ($x_1$, $x_2$, ..., $x_n$, $y_1$, ..., $y_m$, $z_1$, ..., $z_l$) as joint distribution p($x_1$, $x_2$, ..., $x_n$, $y_1$, ..., $y_m$, $z_1$, ..., $z_l$). Each node of the graph represents a random variable, while edges represent the conditional dependence or correlation between the variables. In particular, we determine the weights ($w$) and biases ($b$) characterizing the connection during a training phase, and then we use them for prediction in an inference phase. Internally, BCPNN is built up using hypercolumns (HCUs), which correspond to a particular variable. For example, if trying to identify numbers, a particular HCU might learn how to capture the number '5'. Inside HCUs are minicolumn units (MCUs), which capture a particular instance of the variable. For example, inside an HCU that captures the number of '5', each MCU might learn a different version of the number '5' (one rotated, one skewed, etc.). The BCPNN network capacity is thus a function of both how many HCUs the network has as well as how many MCUs are inside each HCU.

Unlike traditional DL, which relies on backpropagation for training the network, we use a localized (and unsupervised) brain-like rule to determine the neural network's weights and biases. In our approach, the learning of the graph connection weights complies with Hebb’s postulate: learning only depends on the available local information provided by the activities of the pre- and post-synaptic units. Instead, back-propagation learning requires gradient signals to be communicated from distant output layers. A Hebbian learning rule allows higher scalability and better utilization of HPC systems.

\begin{figure}[t]
    \centering
    \includegraphics[width=0.98\linewidth]{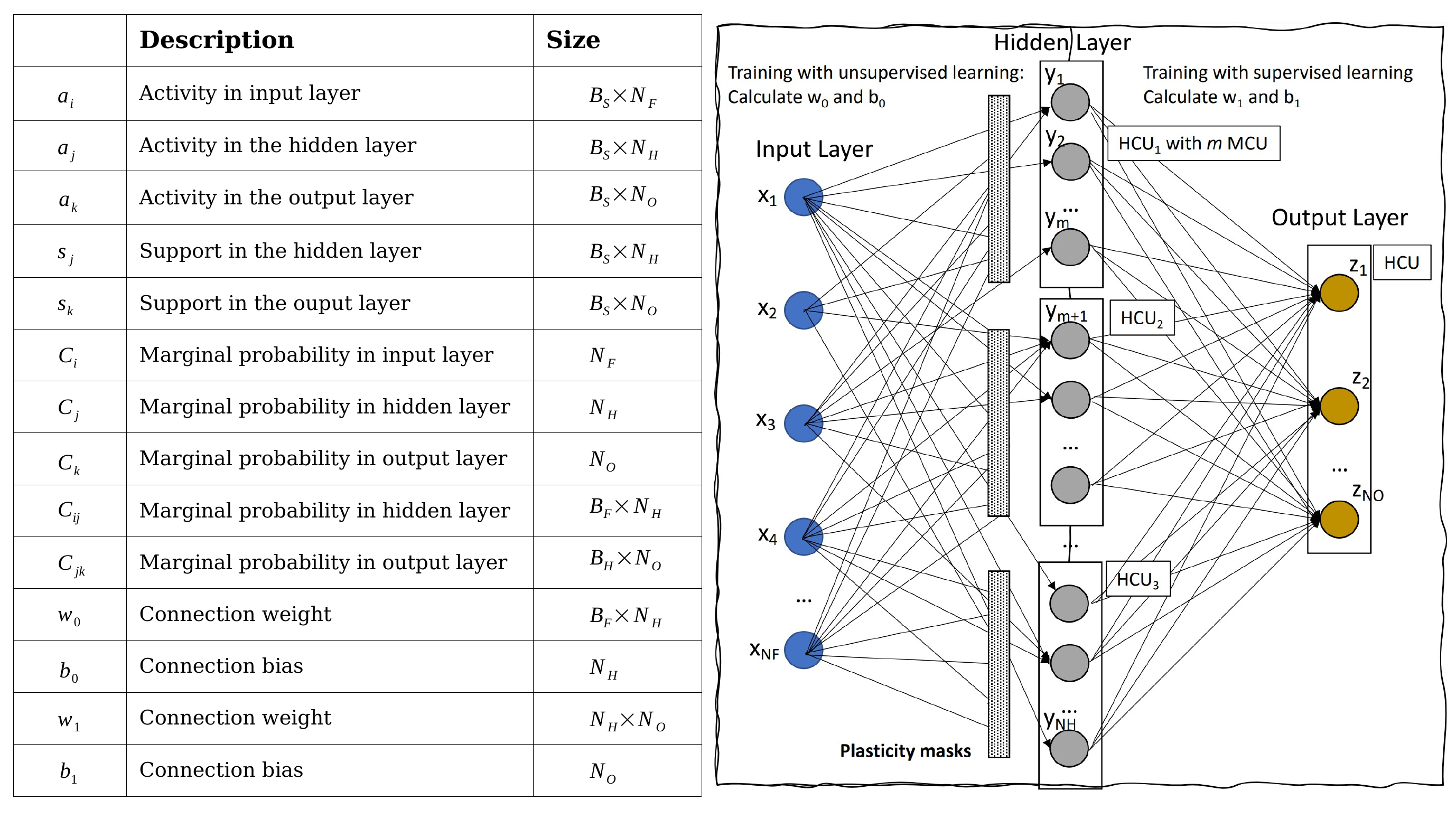}
    \vspace{-0.5cm}
    \caption{Array and matrix sizes for different quantities associated to the BCPNN graphical model.}
    \label{fig:network}
    \vspace{-0.7cm}
\end{figure}

For the sake of simplicity, we consider a graphical network with three layers: input, hidden, and output layers. A difference with traditional neural networks is the organization of the hidden layer as HCUs and MCUs to structure the local activity. A second innovation is the dynamics of change of the sparse connectivity between the input and hidden layer. This is achieved by using masks that silences connections from the input layer to the hidden layer. A third feature is the dynamic regulation of unit biases.  We show an example of such a neural network in Fig.~\ref{fig:network}. For the equations regulating the learning and bias control, we refer to the paper by Ravichandran et al.\cite{ravichandran2020learning}. In this paper, we focus on the algorithm emphasizing the computational cost and the good fit to HPC systems.

We summarize different quantities related to the BCPNN that are used for the training and inference in the table in Figure~\ref{fig:network}. The subscripts $i, j$, and $k$ refer to quantities in the input, hidden, and output layer,  respectively. We also report the sizes of arrays and matrices that will determine the computational cost. $N_F$ is the number of input features, $N_H$ is the number of the hidden layer units, and $N_O$ is the number of output layer units, e.g., the number of classes in a classification problem. Unlike previous BCPNN implementations, we introduce $N_B$ mini-batches of size $B_S$ for training the network. 

The training of StreamBrain consists of two steps: we first have an unsupervised step where we calculate the weights and biases ($w_0$ and $b_0$) connecting the input layer to the hidden layer. After that, we freeze these connections and then perform supervised training to calculate the connection weights and biases ($w_1$ and $b_1$) between the hidden layer and output layer. The algorithm for the training of the hidden layer is shown in Algorithm~\ref{al:train-1} (training of the output layer is similar, and is not shown for conciseness).

\setlength{\textfloatsep}{0pt}

\begin{algorithm}[t]
    \caption{\label{al:train-1}Unsupervised training of hidden layers }
    \small
    \KwResult{Calculate $w_0$ and $b_0$}
    \For{$ 1$ \textbf{to} $n_{epochs1}$}{
        Shuffle input data\;
        \For{$ i_B \gets 1$ \textbf{to} $N_{B}$}{
            \If{$i_B \% N_{HCU}== 0$}{
                Update plasticity mask\; 
            }
            $a_i \gets $ batch of input data\;
            $s_j \gets  a_i w_0 + b_0$\; 
            $a_j \gets $ softmax($s_j$) over the HCU\;
            \For{$1$ \textbf{to} $n_{cycles}$} {
                $C_i \gets (1 - \lambda) C_i + \lambda \langle a_i \rangle$\;
                $C_j \gets (1 - \lambda) C_j + \lambda \langle a_j \rangle$\;
                $C_{ij} \gets (1 - \lambda) C_{ij} + \lambda \langle a_i \tens{} a_j \rangle$\;
                $w_0  \gets \log (C_{ij} /  C_i \tens{} C_j)$\;
                $b_0 \gets k_B \log C_j $\;
                Apply mask to $w_0$;
            }
        }
    }
\end{algorithm}

The parameter $\lambda$ is the exponent of an exponentially-weighted average, and is a key parameter that shapes learning dynamics. The $\langle ... \rangle$ indicates an average over a single batch consisting of $B_S$ samples. In the BCPNN, the hidden layer is organized as a series of HCUs with several MCUs. The activation function is calculated within an HCU with a softmax operation (see line 9 in Algorithm~\ref{al:train-1}): $a_j = \exp(s_j) / \sum_m  \exp(s_m).$

A major innovation is the introduction of structural (dynamic) plasticity. This is implemented by using masks that are applied to the weight $w_0$ matrix as element-by-element matrix multiply. Initially, we randomly set the plasticity (and also the mask), and then it evolves depending on the correlation between unit activities. To calculate the mask we compute the mutual information for connections between the input and hidden layer. Since the total number of active incoming connections is fixed, each HCU greedily maximizes the mutual information it receives by silencing the active connection with the lowest mutual information and activating the silent connection with the highest. The inference step is similar to the traditional DL approach.

\subsection{BCPNN Performance Model}
The main computational kernel used in the training and inference step of the BCPNN network is the batched outer product of arrays. The outer product of the two arrays is calculated as dense matrix-matrix multiplications. By investigating Algorithm~\ref{al:train-1}, neglecting the cost of calculation of structural plasticity and assuming $N_H >>  N_I >> N_O$ and $B_S >> 1$ as in the majority of use cases, the computational cost for training is dominated by batched $B_S$ matrix multiplies in lines 8, 13 of ~\ref{al:train-1},  resulting in an upper bound cost of: $T = \mathcal{O}(n_{cycles} * N_B * B_S * N_H * (N_F + N_O))$. The computational cost scales linearly with the number of units in the hidden layer $N_H$. The batch size of $B_S$ increases the computational workload, thus increasing the usage of CPUs and accelerators. 


\section{StreamBrain Design}\label{sec:streambrain}
StreamBrain is our attempt at streamlining the use of emerging brain-like neural networks machine learning models, and for further use in emerging HPC and data-center use-cases. We aspire to have StreamBrain support two primary modes of operation, which would cater to different needs for different users. These two operations are: \textbf{(i)} Streaming, which allows a third party (e.g., a network card or camera) to deliver input data to the application at variable (and unpredictable) latencies, which are used to either train or infer using a network, \textbf{(ii)} Batched, which is similar to existing DL frameworks (but also incorporates the notion of time), for both training and inference. This paper focuses exclusively on exploring and investigating the batched execution mode. StreamBrain as a framework is implemented in Python, albeit core parts -- particularly those that are computationally heavy -- have been factored out and optimized with OpenMP, CUDA, or FPGA backends. A prototype version of StreamBrain has been made available at~\footnote{https://github.com/KTH-HPC/StreamBrain}.


\begin{figure}
\begin{lstlisting}[language=Python,caption=Describing a BCPNN Network in StreamBrain,label=lst:bcpnn-streambrain-code]
# 1. Create empty network
model = BCPNN.Network(...) 
# 2. Add layers
model.add(BCPNN.StructuralPlasticityLayer(..)) 
model.add(BCPNN.DenseLayer(...))
# 3. train and evaluate
model.fit(dataset=(...)) 
model.evaluate(dataset=(...))
\end{lstlisting}
\end{figure}

\textbf{StreamBrain DSL}: To facilitate an easy, portable, and familiar interface for using BCPNN, we created a Keras-like interface for StreamBrain. In this paper, we focus primarily on three-layer BCPNN use-cases that combine unsupervised (hidden layer) and supervised (output layer) training, which is implemented in our DSL using few lines of code (see Listing~\ref{lst:bcpnn-streambrain-code}). StreamBrain is described in Python, allowing using it standalone or integration into existing ML pipelines. We also support multiple HPC backends.

\textbf{StreamBrain Backends}: The StreamBrain implementation is based on using the Python Numpy module and expressing the operation in Algorithm 1 as arrays and tensor operations. We implemented a series of backends to support accelerators targeting CPU, GPU, FPGA, and MPI. The backends can be switched easily through an environment variable and they are interfaced with the StreamBrain framework through Python bindings.



\textbf{Python Implementation:} StreamBrain is implemented in Python, leveraging NumPy where applicable for performance reasons. We have structured the implementation according to the function they perform, according to Algorithm~\ref{al:train-1}. The first group of methods is responsible for computing activations inside the network (~\ref{al:train-1}:L7-9). The second group of methods is responsible for learning and updating weights (Algorithm~\ref{al:train-1}:L10-16). The final group is responsible for updating the receptive fields on layers with structural plasticity (Algorithm~\ref{al:train-1}:L4-6). Out of all the methods implementing, perhaps the most important ones are the  \texttt{updateMarginals()}(L11-L13 in Algorithm~\ref{al:train-1}), the \texttt{updateWeights()} and \texttt{updateBias()} (L14 and L15 in Algorithm~\ref{al:train-1}). These are the most computationally expensive, relying on an optimized BLAS library for performance; they are also the most salient candidates for accelerations. The structural plasticity  (Algorithm~\ref{al:train-1}:L4-6) follows the description in~\cite{ravichandran2020learning}, and computes a score for each position in the receptive field and silence active connections with the lowest score while activating connections with the highest score. 

\textbf{CPU Backend:} Our CPU-backend use OpenMP primitives to parallelize those kernels that are the main computational bottlenecks (obtained through profiling) in the BCPNN model. These computational bottlenecks are mainly associated with the Python functions \texttt{updateMarginals()} as well as the \texttt{updateWeights()}. We merged the updates on $C_i$ and $C_{ij}$ to maximize temporal locality, and used OpenMP data-parallelism (\texttt{\#pragma omp parallel for}) to distribute work across worker threads. Inside threads, we manually inserted vector operations to further increase performance. For matrix multiplications (such as the inference step, Algorithm~\ref{al:train-1}:L8), we called Intel MKL's S/DGEMM function.

\textbf{GPU Backend:} Our GPU backend was designed to fully run BCPNN on the GPU with little interaction with the orchestrating host CPU. We implemented it using CUDA and leveraged cuBLAS, adapting multiple techniques, including local prefetching, blocking, and preprocessing. Each CUDA warp operates on a single HCU, which allows us to use the \texttt{warp shuffle} functionality to finding sums of all elements inside an HCU in an efficient manner (for the softmax and structural plasticity). We have found this solution to work well, albeit for a much smaller (or larger) number of MCUs, a different implementation might make better use of the hardware. The structural plasticity mask is updated by a single warp, which will score the connections and activate/silence new connections; the infrequent updating of the structural plasticity makes it not the primary candidate for performance optimization. See our open-source implementation for more details.

\textbf{FPGA Backend:} To support emerging HPC and data-center infrastructure, we explicitly added support to offload parts of the computation to Field-Programmable Gate Arrays (FPGAs), currently targeting the Stratix V DE5-Net board. Rather than describing the hardware using low-level (and less portable) Hardware Description Language (HDLs) such as Verilog or VHDL, we use High-Level Synthesis (HLS). We used Intel OpenCL SDK for FPGA~\cite{czajkowski2012opencl}, which was primarily driven by choice of hardware, our prior experiences~\cite{zohouri2018high,podobas2017designing} and to accommodate support for the upcoming Intel OneAPI. 

For the FPGA implementation, we focused on the partial acceleration of the two most heavy components of the BCPNN models: \texttt{updateMarginals()} and \texttt{updateWeights()}, both of which we merged into a single FPGA kernel in order to preserve space, encourage the sharing of resources and increase temporal data locality. Several components make up our FPGA accelerator. An address-generator is responsible for prefetching most of the data from external DDR memory and storing the fetched data in local blockRAM (a memory resource unique to FPGAs). A custom matrix engine will perform the necessary BLAS-3 matrix-matrix operation on the fetched data, before streaming the result to the network probability unit, which finalizes the update and streams the data back to external DDR memory. In this paper, we consider most of the data to be stored in external (DDR) memory, which can (in future work) trivially be extended to handle a stream of data (e.g., from a camera) for use with external devices. Unlike CPUs and GPUs, which work with a predetermined fixed floating-point precision format, our StreamBrain accelerator can vary the type of floating-point representation that is used. More specifically, all \texttt{additions}, \texttt{subtractions}, \texttt{multiplications}, \texttt{division}, and \texttt{logarithm} floating-point functions can be varied, which is a property we exploit to study the resilience to numerical precision that BCPNN has (not only multiply-accumulate as in NVIDIA Tensorcore or Google TPU). Such studies are also imperative to later guide a BCPNN ASIC accelerator. We implemented variable precision through custom Register Transfer Level (RTL) VHDL code generated by FloPoCo~\cite{de2011designing}, and created a custom OpenCL library with these variable-precision operators inside, allowing said operators to be invoked through regular C-like function calls. We investigate variations of the IEEE-754 single-precision, but with reduced mantissa, and can call them BF28 down-to BF14-- our BF16 representation is identical to the one in e.g., Google TPUs.

\textbf{MPI Backend:} To support training with large-scale datasets across multiple nodes on HPC systems, we extend the CPU Backends to use data-parallelism through a hybrid MPI+OpenMP approach. In each step, a batch is further divided by the number of processes where each process is responsible for a sub-batch. Weight updates are performed using \texttt{MPI\_Allreduce()} to derive a global mean operation over all the batches (for L11-13 in Algorithm~\ref{al:train-1}) before updating the marginal probabilities ($C$) and weights ($w$) locally. For all other CPU Backends, we distribute work using block distribution and finalize using \texttt{MPI\_Allgatherv()}. Inside each process, OpenMP is used to further parallelize the computation.

\section{Results}\label{sec:result}

We evaluated StreamBrain on a broad and diverse set of architecture, including GPUs and FPGAs. The systems were as following:

\noindent {\bf 1) \beskow{}} is a supercomputer at KTH based on the Cray XC40, where each node has two Xeon E5-2698v3 Haswell 2.3 GHz, 64 GB RAM, running Python 3.7 and MKL+IntelMPI+ICC 19.0.1.144,

\noindent {\bf 2) \kebne{}} is a supercomputer at HPC2N, containing either two Intel Xeon E5-2690v4 or Gold 6132 processors per node, which also contains NVIDIA Volta-100 GPUs (PCIe), running OpenMPI 3.1.3, GCC 8.2.0, Python 3.7.2, MKL 2019.1.144, and  CUDA 10.1.243,

\noindent {\bf 3) \textbf{A100-system}} is local KTH node with a AMD Epyc 7302P (16-Core) processor, a NVIDIA Ampere-100 GPU (PCIe), running GCC 8.3.1, CUDA 11.1, Intel MKL, and Python 3.8,

\noindent {\bf 4) \textbf{FPGA-system}} is a local KTH node with an Intel Core i5-8400 with an Intel DE5-Net board (Stratix V 5SGXEA7N2F45C2).

All evaluations applied aggressive optimizations (\texttt{-O3}). All GPU experiments were done with a single GPU (NVIDIA A100 or V100). For the FPGA evaluation we disabled caching (\texttt{-nocaching}), relaxed floating-point ordering (\texttt{-fp-relaxed}), and created custom IEEE-754 derived FPUs (for addition/subtraction/multiplication/logarithm) using FLoPoCo~\cite{de2011designing} and manually assembled them into an OpenCL library. We enabled the \texttt{OMP\_PROC\_BIND=true} environment variables for those platforms that gained from using it. NumPy was recompiled with support for Intel MKL. For the evaluation, we used MNIST~\cite{lecun1998mnist} and STL-10~\cite{coates2011analysis}, both well-known and well-used image recognition benchmarks. For both the MNIST and STL-10 experiments, we use a 3-layer network (input, hidden, output) where the hidden layer is composed of 3000 MCUs and the number of HCUs a hyperparameter. The batch-size varies depending on the experiment. We optimize hyperparameters by doing 1000 runs as follows: \textbf{(i)} 250 quasi-random samples using the AX platform (\url{https://ax.dev}), \textbf{(ii)} one worker selects a hyperparameter using the GPEI algorithm from the AX platform, and \textbf{(iii)} all other workers select hyperparameters using TBPSA algorithm from Nevergrad.

\subsection{StreamBrain CPU and GPU performance}
We start by quantitatively evaluating StreamBrain on both general-purpose processings (CPUs) and Graphics Processor Units (GPUs) using the well-known MNIST~\cite{lecun1998mnist} handwritten digit recognition benchmark. We use MNIST (instead of, e.g., the larger ImageNET or STL-10) because the performance (in terms of accuracy) of MNIST using BCPNN is well-understood and documented, allowing us to examine compute performance of StreamBrain and correctness. 

\begin{figure*}[t]
	\begin{center}
		\includegraphics[trim=0 7.5cm 0 0.5cm, width=2.15\columnwidth]{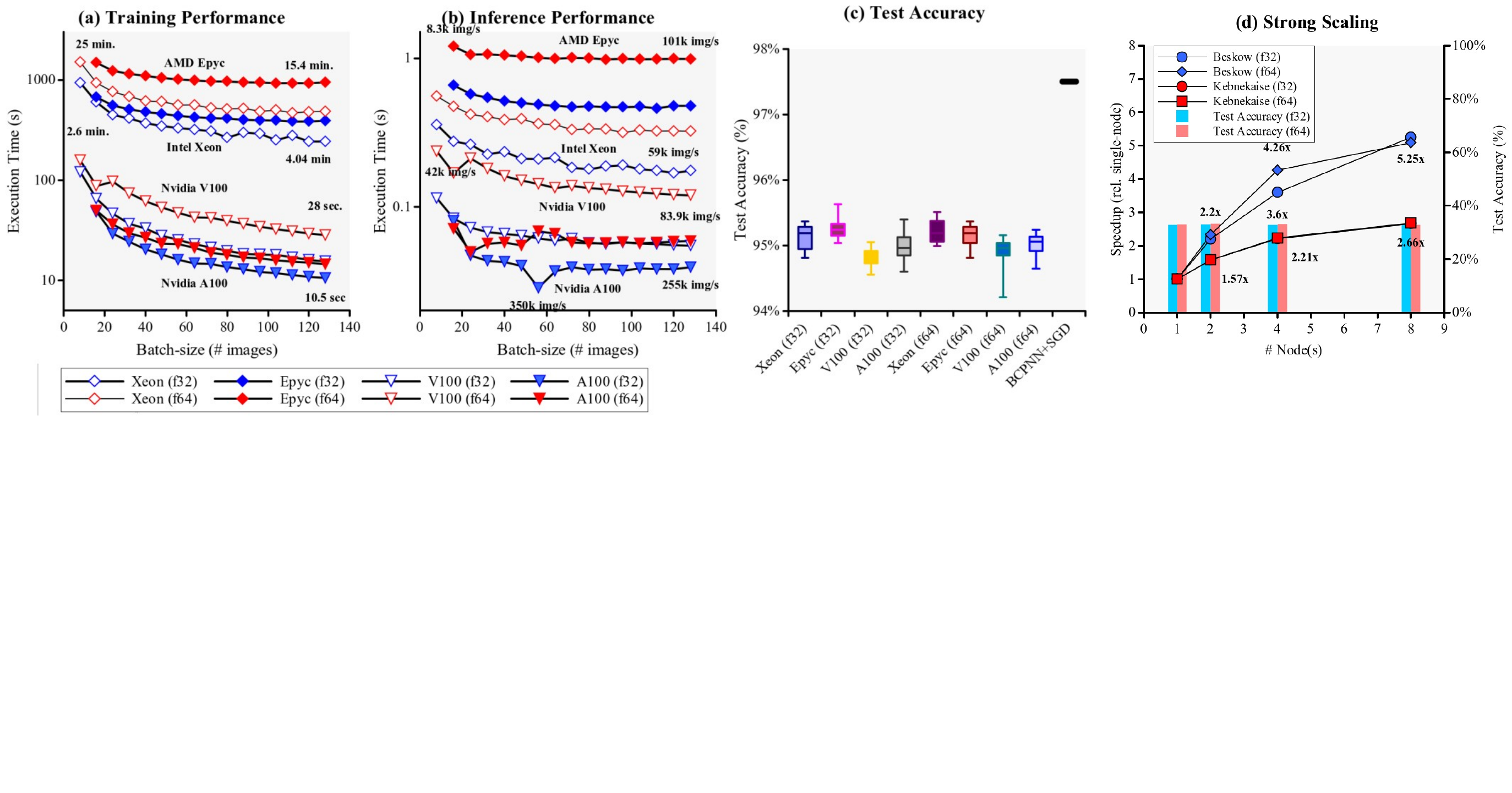}
		\caption{StreamBrain performance with a number of general-purpose processors (CPUs) and graphics procesing units (GPUs), including \textbf{(a)} training performance, \textbf{(b)} inference time, and \textbf{(c)} test accuracy for the MNIST benchmark, and (d) strong scaling on HPC-systems using the STL-10 benchmark.}
		\label{fig:sbtrain}
	\end{center}
	\vspace{-.5cm}
\end{figure*}

Fig.~\ref{fig:sbtrain}:a shows the performance of StreamBrain as a function of batch-size on MNIST for both CPUs and GPUs, using both IEEE 754 single- and double-precision (named f32 and f64 respectively). Overall, we notice that the performance increases as a function of batch-size; a larger batch-sized encourage more data-parallelism~\cite{ben2019demystifying} and locality, turning multiple expensive BLAS2 operations into a single BLAS3 operation. We also notice that the difference between the CPU and GPU StreamBrain is rather large, ranging between 7.75x-65x in favor of the GPUs; this difference is larger with double- compared to single-precision, is also because more code remains in Python for the CPU (contra GPU) version. We also see an (expected) near two times increase in performance of decreasing the width of the numerical representation and going from double- to single-precision increase the performance of $\sim2\times$ on the CPUs and between 1.7x and 1.26x (on average) on the GPUs. Training the entire MNIST dataset using StramBrain can be as fast as $\sim10$ seconds using the NVIDIA A100 GPU or $\sim4$ minutes on a server-class Xeon CPU. The (seemingly) large difference between AMD Epyc and Intel Xeon processors is because our nodes have dual-socket Intel Xeons (while we only used a single AMD Epyc processor). Fig.~\ref{fig:sbtrain}:b shows inference (or prediction) performance of the StreamBrain framework. Here, the difference between the CPU and GPU is lower and can be as low as 3x difference (Xeon vs V100 in some cases), albeit on average the CPUs are between 5x-8x slower. As with training, the performance scales with the batch-size: single-image inference (or "streaming") reach between 28k to 87k images/second, while a larger batch-size allows up to 350k images/second to be achieved by the GPUs.  Fig.~\ref{fig:sbtrain}:c shows the inference accuracy, here averaged across all batch-sizes. Overall, all implementations yield an average that is above 95\%, which is roughly ~1\% away from the \textit{boosted learning} in~\cite{ravichandran2020learning}. There are some discrepancies both between using single- and double-precision, as well as between the GPU and CPU versions. The difference in architecture is likely because of the different random generators used to initialize the network at the start. Finally, 97.5\% of accuracy can be reached by using a hybrid solution: using StreamBrain to derive hidden layer representations using unsupervised learning and use a stochastic gradient descent (SDG) training only for the output layer, demonstrating correctness with the hybrid approach as reported in~\cite{ravichandran2020brain} (average of 97.77\%). We end by noting that training a network to reach ~95.5\% with StreamBrain on MNIST is faster (10.5 seconds) than training an MLP of similar capacity with PyTorch (33.94s $\pm$ 1.04 seconds) to reach the same ~95.5\% accuracy on the modern Nvidia A100, showing a benefit in performance over PyTorch (albeit, given longer time, the PyTorch version will reach a better accuracy).

\subsection{StreamBrain FPGA Exploration}

\begin{figure}[t]
\vspace{-0.3cm}
     \begin{center}
        \includegraphics[width=0.4\textwidth]{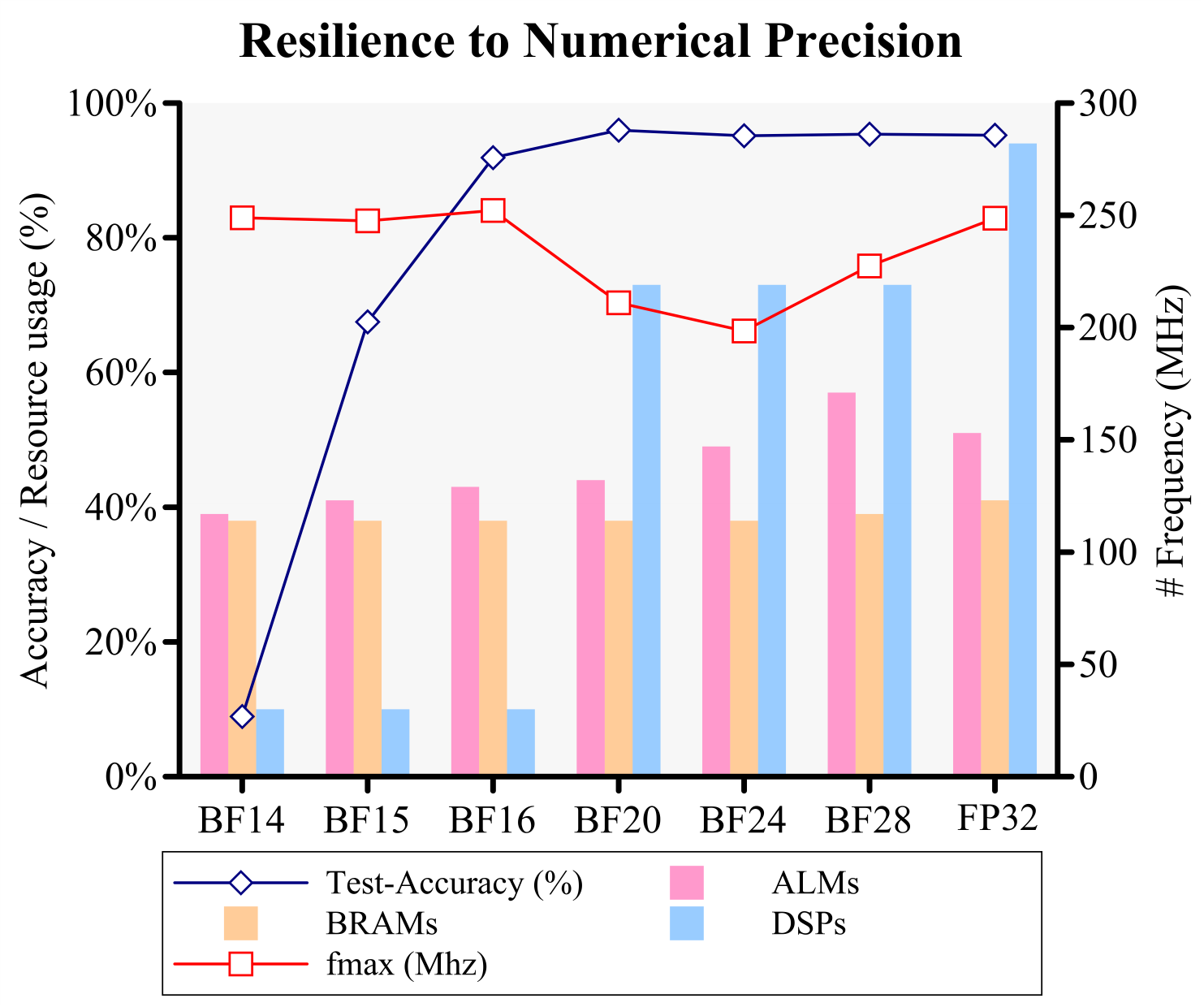}
        \vspace{-0.15cm}        
        \captionof{figure}{Effect of reducing precision on test accuracy and resource usage (blue square and bars, left y-axis) and on frequency (red square, right y-axis)\label{fig:fpga_perf}}
        \end{center}


\end{figure}

In the previous section(s), we evaluated our high-performance BCPNN implementation in StreamBrain on the MNIST benchmark to verify correctness with prior work. We also observed that the testing accuracy degradation between IEEE-754 single- and double-precision was negligible, yielding the follow-up question: how tolerant (or resilient) is BCPNN to reductions in the number representation? A smaller representation can significantly increase performance (more data per unit bandwidth) and reduce the silicon FPU footprint. In this section, we leverage our FPGA implementation to explore number representations that (for performance reasons) are hard to explore in CPUs (simulating different representations is slow). More specifically, we create custom hardware that uses (the today well-known) brain-float 16 (BF16) representation, as well as create alternative versions that we call BF14, BF15, BF20, BF24, and BF28 (that have between 5- and 19-bit mantissa), and accelerate the most compute-intensive functions of StreamBrain on the FPGA. Despite using a rather old FPGA (from 2010), the performance of our FPGA accelerator is faster than the original code in Python/NumPy, and is comparable to that of the CPU versions. Fig.~\ref{fig:fpga_perf} shows the training accuracy (blue diamonds) as a function of numerical representation, and we see that the BCPNN model is resilient enough to tolerate down-to BF16 with minor ($\sim4\%$) accuracy degradation; BF20 and higher experience no accuracy degradation compared to single-precision. BF14, however, drops down to mere chance ($\sim10\%$) for the MNIST dataset, while BF15 is between (67.5\%). While the operating frequency(red squares)  of using a different representation tends to stay roughly the same (between 198 and 252 MHz), the resource utilization tends to decrease as a function of reduced numerical representation. While all types of resources decrease with smaller representation, interestingly, the DSP utilization drop significantly (from $\sim70\%+$ to 10\% when using BF16 and below. We cannot fully understand this jump in reduction, but it is likely that smaller representations are synthesized to logic (rather than DSPs). In short, our experiments show that BCPNN would be a good candidate to accelerate on BF16-capable devices, such as Google's TPU or the upcoming Power10 or Intel Sapphire Rapids. In future work, we would use the FPGA to explore alternative representations (e.g., Posit~\cite{gustafson2017beating,podobas2018hardware}).

\subsection{Higher-Dimensional Problems}
In the previous section (and also in prior work), applying BCPNN was limited to MNIST-sized data-sets primarily due to the lack of a high-performance framework. With StreamBrain, for the first time, BCPNN can now be scaled to more complex high-dimensional problems. We trained the STL-10 dataset ($\sim30+$ times larger than MNIST) using BCPNN for 100 (hidden) and 20 (output) epochs using StreamBrain's GPU backend on the NVIDIA A100 on a network with 3000 MCUs (20 HCUs). To contrast our performance, we also trained an MLP network with similar capacity using PyTorch (AdamW optimizer, cross-entropy loss function, ReLu activations) also for 100 epochs.
The time to train the BCPNN network was 178.2$\pm$0.1 seconds and yielded a test accuracy of 34.8$\pm$4.9\%. The (deep learning type) PyTorch trained network took 100.2$\pm$0.43 seconds and yielded a test accuracy of 42.2$\pm$0.12\%. Ignoring the better accuracy (+7.4\%) that the backpropagation-type network gave, both performances are comparable, where our StreamBrain is $\sim77\%$ slower-- a respectable number given the difference in the number of manhours spent in StreamBrain versus PyTorch.

\subsection{Strong scaling} 
We end the result section by measuring the strong scaling properties of StreamBrain (MPI-backend) when given between one and eight nodes of computing capacity on the \kebne{} and \beskow{} supercomputers. The problem to solve is to train on the STL-10 benchmark under a batch-size of 512. Fig.~\ref{fig:sbtrain}:d shows the speed-up (relative to single-node performance). We see that the speed-up on the \kebne{} machine is very stable, and both the single- and double-precision version reaches (near) identical performance, albeit capping out at 2.7x. For the \beskow{} computer, there is one anomaly when using four nodes between the single- and double-precision version, which we still do not fully understand. At eight nodes, the \beskow{} supercomputer yields a peak of 5.25x speed-up. Note how the test accuracy, even with a batch-size of 512, remains comparable to that with lower batch-sizes (previous section).

\section{Related Work}\label{sec:related}
Our contribution, StreamBrain, has been primarily inspired by the many different DSLs and libraries that exist for use in deep learning. We name Keras~\cite{gulli2017deep} as our primary source of inspiration, but equally intuitive interface also exists in PyTorch~\cite{paszke2019pytorch}. Outside of DL, the literature and availability of such frameworks are sparse. For the neuroscience community, domain-specific mark-up languages and libraries such as Neuron~\cite{hines1997neuron} and NEST~\cite{gewaltig2007nest} provide an interface to simulating highly accurate and exact spiking neural networks. Despite the relatively large user base that these tools have, they are not a good match for applying emerging brain-like models in practice, as they are often too detailed (and hence slow to simulate) and aimed at simulating the biological brain; using these tools requires significant expertise and effort. Moving an abstraction layer up, simulation frameworks such as PyNN~\cite{davison2009pynn} offers a DSL for describing populations of spiking neurons and multiple simulation backends. At this level, it is possible to undertake experiments and evaluations for practical tasks such as the image recognition we do here. However, many of these frameworks still incur a steep learning curve. There is, however, one framework that provides an interface similar to the one of StreamBrain: Nengo~\cite{bekolay2014nengo}. Nengo's Brain Maker allows the simple creation of brain-like neural network models, both spiking and non-spiking, for use in ML, with examples showing, for example, MNIST. Nengo supports the same backends as StreamBrain, including CPU, GPU, and FPGA~\cite{morcos2019nengofpga} implementation. Unlike Nengo, StreamBrain focuses on BCPNN.

\section{Conclusion}\label{sec:conclusion}
We have introduced StreamBrain -- a high-performance DSL targeting the BCPNN model. We presented and evaluated four different backends on GPUs, FPGAs, and CPUs. We show how to train MNIST as fast as \textit{10 seconds}, and showed results (for the first time) on higher dimension problems such as STL-10. We also introduced batching into BCPNN and showed that BCPNN could work with low-precision arithmetic. Our contribution enables future exploration of BCPNN in HPC Computing.

\noindent{{\textbf {\footnotesize Acknowledgements: }}\scriptsize This work was funded by the European Commission H2020 program under grant agreement no. 801039 (EPiGRAM-HS) and  no. 800999 (SAGE2). This work was also supported by the Swedish e-Science Research Centre (SeRC). The computations were enabled by resources provided by the Swedish National Infrastructure for Computing (SNIC) at PDC and HPC2N and partially funded by the Swedish Research Council through grant agreement no. 2018-0597.}

\bibliographystyle{ACM-Reference-Format}
\bibliography{ref}

\end{document}